# Three limits to the physical world

*Pierre Darriulat*
*Hanoi University of Sciences and VATLY/INST,*
*179 Hoang Quoc Viet, Cau Giay, Hanoi, Vietnam*
darriulat@mail.vaec.gov.vn

We present below a simple diagram (Figure 1) that summarizes much of our present knowledge of the physical world. Its substance is very elementary common knowledge, most of which does not require quoting references. Its interest is in the form, not in the substance. As we are dealing with typically hundred orders of magnitude, we shall be satisfied with a precision of one order of magnitude.

In a preceding article [1] an early version had been proposed, using size ($l$) and mass ($m$) as axes. The present version uses instead two new variables, $\xi=l/m$ and $\eta=\sqrt{(lm)}$. As the diagram uses a *log-log* scale, this corresponds essentially to a 45° rotation of the axes. Units such that $\hbar=c=G=1$ are used throughout ($\hbar$ is the Planck constant, $c$ the velocity of light in vacuum and $G$ Newton's gravity constant). Note that $m=\eta/\sqrt{\xi}$ and $l=\eta\sqrt{\xi}$, density being $\rho=(\xi\eta)^{-2}$.

The $\xi$ axis is called "Heisenberg limit" in reference to Heisenberg uncertainty relations. Taking the proton as an example ($\xi\sim10^{39}$), the quark momenta give the scale of the proton mass, $m$, and must exceed the reciprocal of the extension $l$ of their wave packets which gives the scale of the proton size. Hence $lm$ must exceed unity, and $\eta\sim1$ corresponds indeed to the Heisenberg limit. The inclusion of elementary fields and pointlike objects in the diagram would imply accepting some convention, such as interpreting $l$ as the Compton wavelength.

The $\eta$ axis corresponds instead to the "Schwarzschild limit" in reference to the black hole singularity in the Schwarzschild metric, $m\sim l$ or $\xi\sim1$ (precisely, the Schwarzschild radius of an object of mass $m$ is $2m$). Known black holes are indeed found on this axis, stellar black holes (such as Cyg X, $\eta\sim10^{39}$) as well as galactic black holes (such as Sgr A*, $\eta\sim10^{45}$ and Cyg A, $\eta\sim10^{49}$). The Universe ($\eta\sim10^{60}$) is also located on this axis because it is flat and its expansion velocity, which is therefore equal to the escape velocity, reaches light velocity on its horizon. It does not mean that the Universe is a black hole: to be a black hole, an object having $\xi\sim1$ must be isolated in space for Schwarzschild metric to apply.

At the origin, the intersection between the two above axes corresponds to the Planck scale ($\xi=\eta=1$) where quantum physics and gravity are known to be incompatible in their present versions and where most of the current theoretical activity is taking place with extra dimensions, supersymmetry and strings.

As the diagram is in *log-log* scale, lines of equal densities are straight lines making angles of 45° with the axes ($\xi\eta=1/\rho^2$). Examples are nuclear matter density associated with neutron stars, $\rho_{NS}\sim10^{-78}$; degenerated Fermi electron gas density associated with white dwarfs, $\rho_{WD}\sim10^{-87}$; and atomic matter density, $\rho_A\sim10^{-93}$, associated with condensed matter and with stars such as our Sun ($m\sim10^{38}$, $l\sim10^{44}$, $\xi\sim10^6$ and $\eta\sim10^{41}$). Indeed, writing $m_p$ and $m_e$ for the proton and electron masses, $p_F$ for the Fermi momentum and $\alpha$ for the fine structure constant, it is straightforward to see that $\rho_{NS}\sim m_p^4$, $\rho_{WD}\sim m_p m_e^3$ (imposing $p_F\sim m_e/2$ to avoid excessive smearing of the surface of the Fermi sphere and noting that $\rho_{WD}\sim8m_p p_F^3$) and $\rho_A\sim m_p m_e^3 \alpha^3$ (using Bohr radius, $\alpha^{-1}m_e^{-1}$, as atomic scale). Hence $\rho_A \div \rho_{WD} \div \rho_{NS} = \alpha^3 \div 1 \div (m_p/m_e)^3 = 1 \div 10^6 \div 10^{15}$.

Also shown on the diagram is the present density of the Universe, equal to the critical density in the Friedman-Lemaître-Robertson-Walker metric, $\rho_{FLRW}\sim10^{-120}$. The corresponding line is labeled "Einstein limit" in reference to Einstein's cosmological constant: the Universe is known to be currently dark energy dominated, dark energy being well described by a cosmological constant $\Lambda$ contributing a term $\rho_\Lambda=\Lambda/8\pi G$ to the energy density of the Universe ($\rho_\Lambda\sim\rho_{FLRW}$).

The third summit of the triangle made by the Heisenberg, Schwarzschild and Einstein limits corresponds to an object having $\xi\sim10^{60}$ and $\eta\sim1$, namely a mass $m_\Lambda$ such that $\rho_\Lambda\sim m_\Lambda^4$, i.e. $m_\Lambda\sim10^{-30}$ or *10 meV*, in the range currently accepted for the lightest known fermions, neutrinos. It implies that dark energy – or the cosmological constant – sets a lower limit to the mass of composite quantum objects in the range of neutrino masses [2]. Indeed, lines of constant masses are straight lines $\eta=m\sqrt{\xi}$ with $m$ ranging from the neutrino mass, $\sim10^{-30}$, to the mass of the Universe, $\sim10^{60}$, with the unit Planck mass ($\sim10^{19}$ GeV) inbetween. Similarly, lines of constant sizes are straight lines $\eta=l/\sqrt{\xi}$ with $l$ ranging from the unit Planck scale ($\sim10^{-33}$ cm) to the size of the Universe, $\sim10^{60}$, with the scale of the "neutrino" wave packet ($\sim10^{30}$) in the middle.

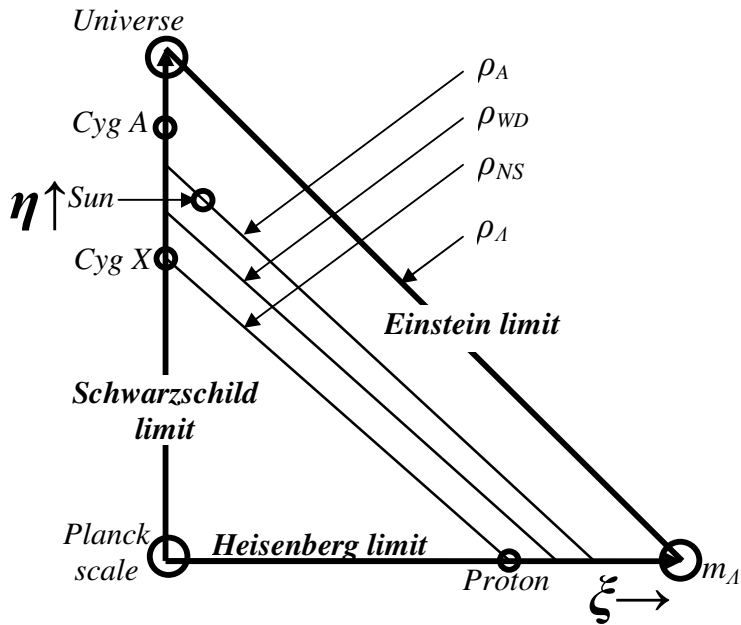

Figure 1. The $\xi$-$\eta$ diagram (see text) in *log-log* scale with $\xi=\eta=1$ at the origin (Planck scale) and $\xi\eta=1/\sqrt{\rho_\Lambda}\sim10^{60}$ on the Einstein limit.

**References**
1. P. Darriulat, Comm. Phys. Vietnam 18/4 (2008) 193.
2. P.S. Wesson, Mod. Phys. Lett. A19 (2004) 1995;
   C.G. Boehmer and T. Harko, Found. of Phys. 38/3 (2008) 2 and references therein.